\pdfoutput=1  

\documentclass{iau} 
\usepackage{graphicx}

\title[Miras in XSL] 
{Oxygen-rich Long Period Variables in the X-Shooter Spectral Library}

\author[A. Lan\c{c}on et al. ]   
{Ariane Lan\c{c}on$^1$,
 Ana\"{\i}s Gonneau$^2$,
 Scott C. Trager$^3$,
 Philippe Prugniel$^4$,
 Anke Arentsen$^5$,
 Yanping Chen$^6$,
 Matthijs Dries$^3$,
 C\'ecile Loup$^1$,
 Mariya Lyubenova$^7$,
 Reynier Peletier$^3$,
 Laure Telliez$^1$,
 Alexandre Vazdekis$^8$
 \and the XSL Collaboration}

\affiliation{$^1$Universit\'e de Strasbourg, CNRS, UMR7550, 
Observatoire astronomique de Strasbourg, \\ 67000 Strasbourg, France -- email: {\tt ariane.lancon@astro.unistra.fr} \\
$^2$ Institute of Astronomy, University of Cambridge, United Kingdom \\ 
$^3$ Kapteyn Astronomical Institute, University of Groningen, The Netherlands \\
$^4$ Centre de Recherche Astrophysique de Lyon (UMR 5574), Universit\'e de Lyon, France\\
$^5$ Leibniz-Institut f\"ur Astrophysik Potsdam, Germany \\
$^6$ New York University in Abu Dhabi, United Arab Emirates \\
$^7$ European Southern Observatory, Garching, Germany \\ 
$^8$ Instituto de Astrofisica de Canarias, Santa Cruz de Tenerife, Spain \\ [\affilskip]}

\pubyear{2018}
\volume{343}  
\jname{Why Galaxies Care about AGB Stars}
\editors{F. Kerschbaum, M. Groenewegen \& H. Olofsson, eds.}
\begin{document}

\maketitle

\begin{abstract}
The X-Shooter Spectral Library (XSL) contains more than 800 spectra of stars across the color-magnitude
diagram, that extend from near-UV to near-IR wavelengths (320-2450 nm). We summarize
properties of the spectra of O-rich Long Period Variables in XSL, such as phase-related features, and 
we confront the data with synthetic spectra based on static and dynamical stellar atmosphere models. 
We discuss successes and remaining discrepancies, keeping in mind the applications to population synthesis
modeling that XSL is designed for.

\end{abstract}

\firstsection 
\section{The X-Shooter Spectral Library  in context}

Modern extragalactic surveys produce energy distributions and spectra of galaxies with an
exquisite precision. Their interpretation calls for progress in the modeling of the integrated light of stellar
populations (e.g. \cite[Powalka et al. 2016]{Powalka16}). With the shift from optical to near-infrared wavelengths that
facilities such as the James Webb Space Telescope (JWST)  or the Extremely Large Telescope (ELT) 
will bring about, it is more important than ever that stellar population synthesis models provide consistent
predictions across the whole spectrum of stellar photospheric emission.

The X-Shooter Library project (XSL, \cite[Chen et al. 2014]{XSL1}) was initiated in this context.
The spectra
extend from the near-ultraviolet (320\,nm) to the near-infrared (2.45\,$\mu$m),
and this avoids some of the inter-connection issues between separate optical and near-infrared empirical spectral libraries that have existed before. Interpolation tools based on XSL will provide empirical spectra at a resolving power 
close to 10\,000 over a broad range of stellar parameters, for direct usage in population synthesis modeling 
(e.g. Verro et al., this volume). In addition, a detailed
comparison of the observed spectra with modern synthetic spectra will help us improve the agreement between the
two, which is vital to validate the fundamental parameters of the stars observed (and hence the way
they can be connected to loci on stellar evolution tracks), as well as to progress towards future population
synthesis models based on synthetic spectra alone. Such purely theoretical models already exist, but the validity of their input synthetic stellar libraries has not been demonstrated over 
a wavelength range as broad as that of XSL.

XSL contains $\sim$800 spectra of more than 650 stars. The second data release, with 
spectra from the three spectral arms of X-Shooter for all program stars, is
nearing completion (Gonneau et al., in prep.). Because luminous cool stars are particularly important contributors to the 
red and near-infrared light of galaxies, XSL was designed to contain a large number of such objects.
The spectra of 35 carbon stars were made available and compared to C-star models 
by Gonneau et al. (2016, 2017).
Here, we  focus on O-rich Long Period Variables (LPVs) with estimated temperatures
lower than about 4200\,K (from Arentsen et al., in prep). The sample used contains 160 spectra of 150 LPVs
of the Milky Way and the Magellanic Clouds, with a range of periods, amplitudes and luminosities.

\section{The spectra of O-rich LPVs in XSL}
\label{sec:AL_description}

The spectra of O-rich LPVs in XSL display a range of colors and spectral features very similar to that
observed by \cite[Lan\c{c}on \& Wood (2000)]{LW2000} [hereafter LW2000], 
which indicates that each of these collections actually 
captures most of the natural variance in the range of colors sampled ($2\leqslant R\!\!-\!\!K \leqslant 10$). 
The XSL spectra have a higher spectral
resolution, especially at optical wavelengths. The existence of two data sets with
such similar properties is useful in the investigation of features that are currently not explained
by models, as it excludes with high confidence that these might be data artefacts.  Such features include 
the molecular emission (most likely of TiO) sometimes seen near 1.24\,$\mu$m 
and the blue slope of the pseudo-continuum around 1\,$\mu$m, both seen preferably in spectra that
also display emission lines and/or strong bands of VO. Note that some of these
properties have been mentioned early in the literature on Miras (e.g. \cite[Wing 1974]{Wing74}).

\begin{figure}[tb]
\begin{center}
 \includegraphics[width=0.47\textwidth]{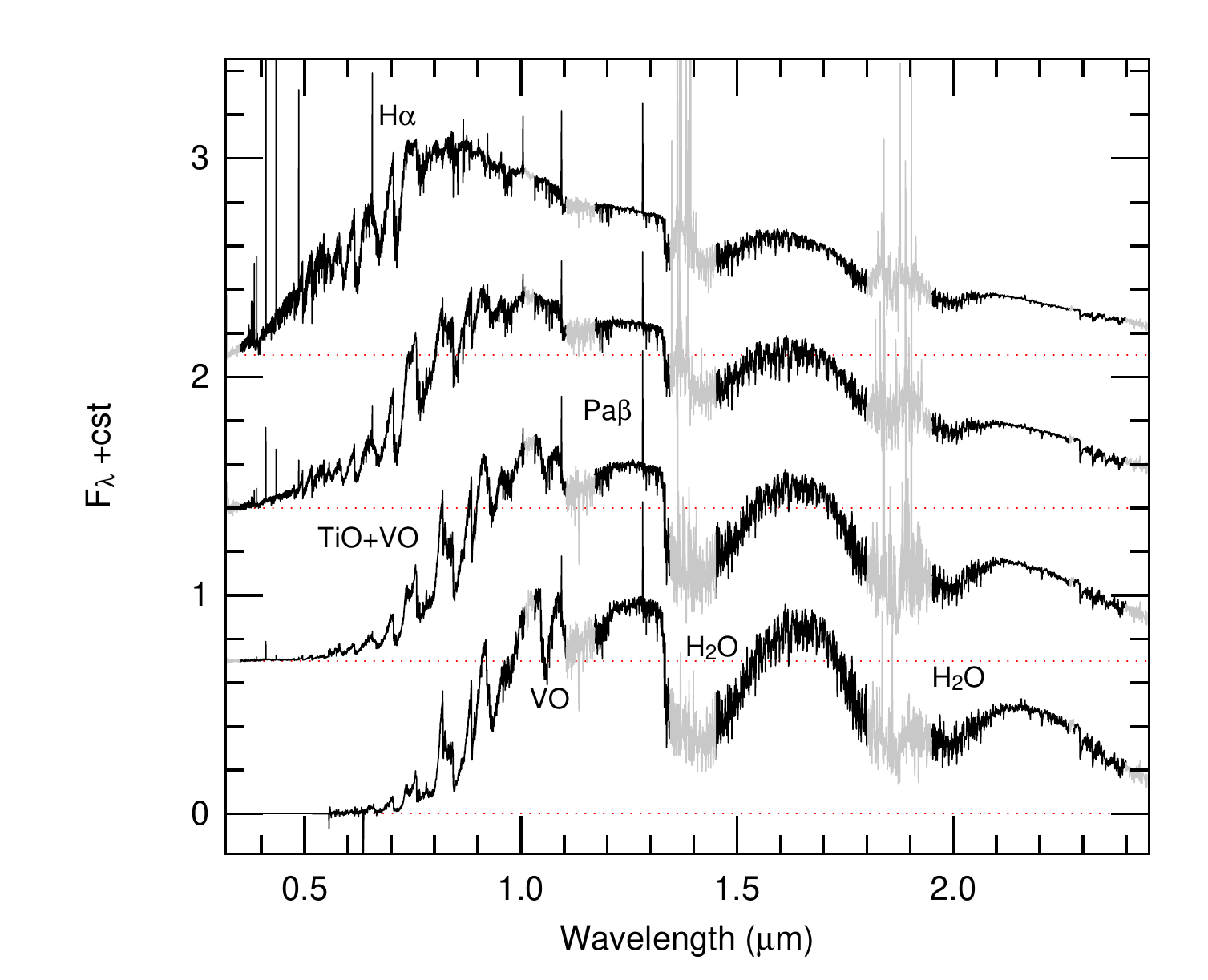}
 \includegraphics[width=0.47\textwidth]{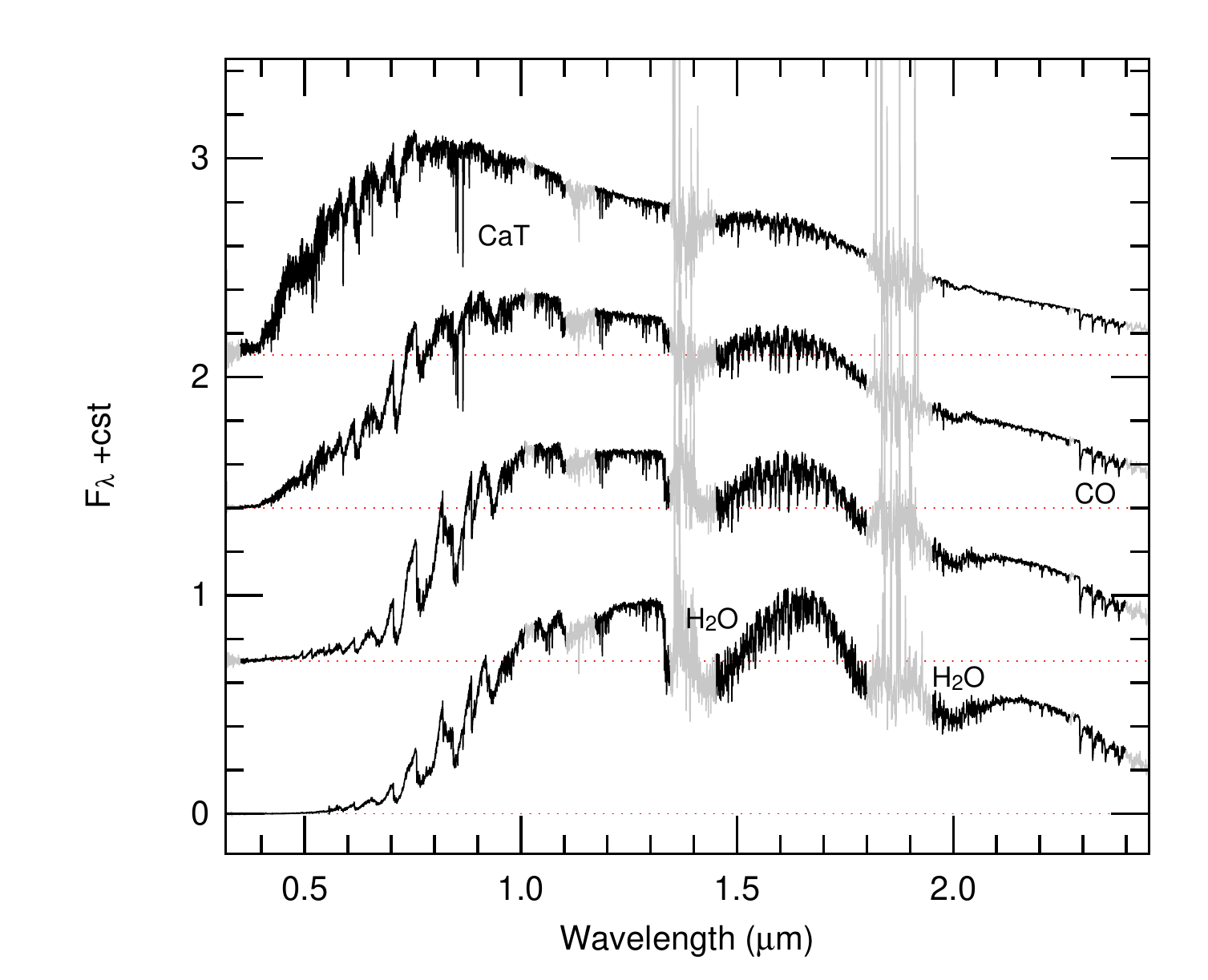}
 \caption{Average spectra of O-rich LPVs from the XSL project, in bins of similar
 $(R-K)$ colors (smoothed to R=3000 for display). Left: spectra at phases that display emission lines. Right: 
 spectra without emission lines. The color-bin definitions are the same in the
 two panels. Grey identifies ranges where the correction for telluric
 absorption leaves strong residuals.}
   \label{fig:AL_spectra_EmL_noEmL}
\end{center}
\end{figure}
 
When averaging XSL spectra in bins of similar broad-band color, as was done earlier
with the LW2000 sample (\cite[Lan\c{c}on \& Mouhcine 2002]{LM2002}), a very regular sequence 
of mean spectra is obtained. This will be useful for population synthesis purposes, but it
hides the real variety of properties, and with the XSL sample we can now also attempt to
exploit this variety. An example is shown in Fig.\,\ref{fig:AL_spectra_EmL_noEmL}, where the left panel 
shows averages of spectra of similar $(R-K)$ color at pulsation phases that display emission lines (typically, but
not exclusively, phases near maximum light), while the right panel shows
averages of spectra in exactly the same ranges of $(R-K)$ color, but selected not to
display emission lines. The energy distributions differ, with more ``triangular" shapes in 
the left panel; the strengths of molecular bands (TiO, VO, H$_2$O)  are stronger in
the left panel, while CO absorption bandheads are more conspicuous in the right panel. 
The calcium triplet lines near 0.85\,$\mu$m are strong in absorption in the right panel, while they 
disappear in the left panel (at the displayed resolution) 
as a consequence of stronger molecular absorption and
line emission. The more rounded shapes in the redder stars of the right hand 
panel might indicate a stronger effect of extinction.

\section{Comparison with models}

At this point in time, we have compared the XSL spectra with two collections of synthetic spectra: (i) the
static PHOENIX models of \cite[Husser et al. (2013)]{TOH2013}, version 2, with solar-scaled abundances and 
a range of [Fe/H]; and (ii) the time series 
of DARWIN models at solar metallicity (Bladh et al., 2015 and this volume). In total, about 2000 model spectra
were considered. But these still represent a tiny fraction of the relevant space of stellar parameters, e.g. in terms
of chemical composition and mass, and for the dynamical models in terms of pulsation properties. All models are
spherically symmetric.

\begin{figure}[htb]
\begin{center}
 \includegraphics[width=0.45\textwidth]{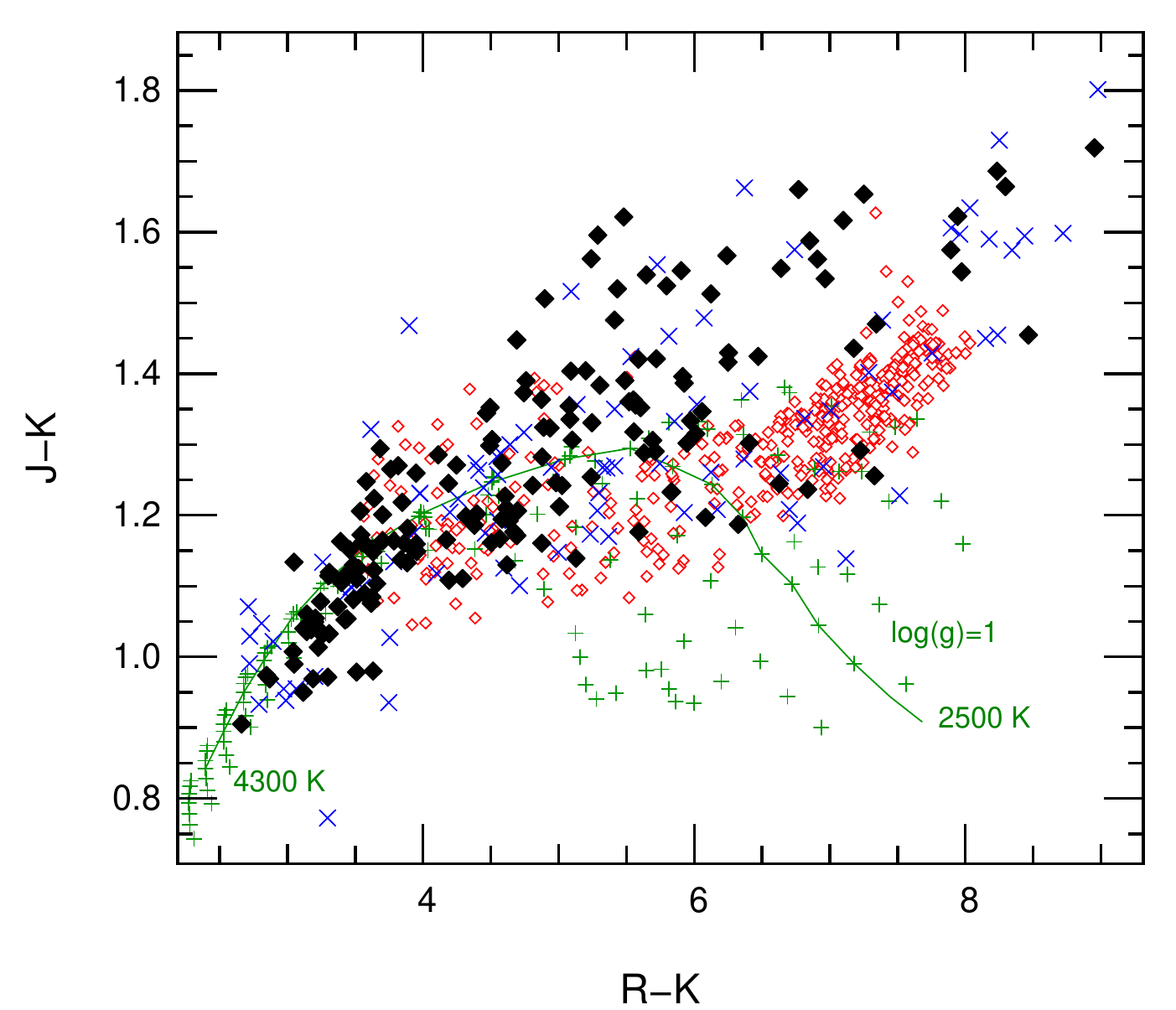}
 \includegraphics[width=0.45\textwidth]{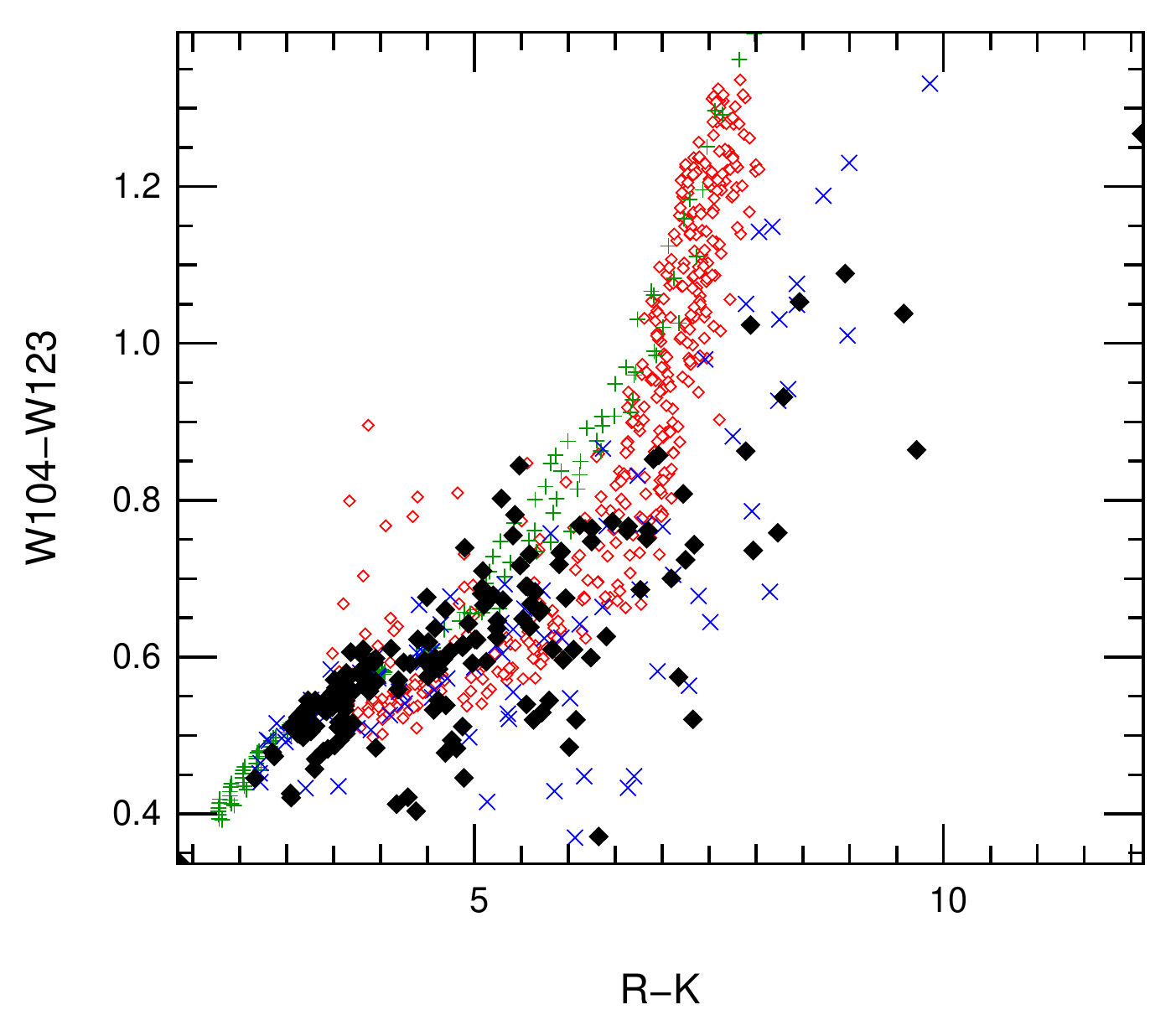}
 \caption{Photometric properties of the observations of O-rich LPVs
 (solid diamonds: XSL; crosses: LW2000), of the Phoenix models at solar metallicity (small plus-symbols; a line connects 
 models at log(g)=1 in the left panel), and of the DARWIN models (small open diamonds).}
   \label{fig:AL_colcol}
\end{center}
\end{figure}

In traditional broad-band color-color diagrams, the loci of the observations tend to look compatible with
the model collections, when allowing for extinction (Fig.\,\ref{fig:AL_colcol}, left panel). However, there are discrepancies in a 
number of diagrams that combine a broad-band color with narrow-band color-indices that target some
of the features mentioned in Sect.\,\ref{sec:AL_description}. For instance, when narrow-band filters
at 1.04 and 1.23\,$\mu$m are used to measure the slope near 1\,$\mu$m, and are plotted against 
$(R-K)$, it is seen that the models do not reach the blue (1.04-1.23) colors observed for LPVs
at some phases (Fig.\,\ref{fig:AL_colcol}, right panel). 
The models also fail to reproduce the strong VO bands often associated with this 
property of the energy distribution (not shown). In these diagnostic diagrams, the dynamical models are displaced
in the desired direction from the static models, but not quite enough. A few dynamical models display molecular bands in emission, but those of CN or CO rather than those of TiO. 

When comparing spectra with models, excellent matches are readily obtained at temperatures near 
4000\,K. For small amplitude variables at those temperatures, a fit to the optical spectrum only, with
a static model and a standard extinction law, often produces a very good representation of the whole
XSL spectrum, with a reasonable match to the detailed spectral features (perfect matches of the metal line
spectra are impossible with only solar-scaled abundances). Below 4000\,K, one cannot expect a good
representation of the near-IR spectrum when fitting only optical wavelengths. Decent representations
of the energy distributions and the main molecular bands (except water and bands near 1\,$\mu$m)
may be obtained when constraining the fit with all available wavelengths. But for spectra near maximum light
the effective temperatures thus obtained with static and dynamical models differ systematically, 
sometimes by more than 300\,K. The most obvious 
discrepancies are found for the strongly peaked spectra that enter the averages in the left panels of Fig.\,\ref{fig:AL_spectra_EmL_noEmL}. The models we have explored until now have a more ``rounded" shape.

\section{Conclusion}

From this early work on the XSL spectra of O-rich LPVs, we have already learnt a few lessons. 
Without surprise, it remains difficult to match the energy distributions and spectral features of 
O-rich LPVs with synthetic spectra. Static models are impressively good matches to small
amplitude LPV spectra at temperatures down to about 4000\,K. In several aspects, dynamical models seem to provide improvements over static models for cool, larger amplitude LPVs
(circumstellar dust, strength of the VO bands, shape of the SED near 1\,$\mu$m), but the range
of properties observed is not covered yet. It is clearly necessary to 
explore a wider range of model parameters in spherical symmetry, and to
start considering 3D models (Liljengren, Freytag, Chiavassa, this volume).

In population synthesis models that predict optical and near-infrared spectra, the warmest
LPVs are most relevant (others have smaller contribution to the integrated light of all stars). 
Average spectra of LPVs remain a convenient practical choice to represent stars on the 
thermally pulsing AGB (TP-AGB). But we have identified the risk of a bias: the bluest stars in
an empirical collection tend to include a larger fraction of stars caught near maximum
light, and at a given broad band color the detailed spectra depend on phase. We hope that
future dynamical models will help us obtain sufficiently good matches so we can relate instantaneous
spectral properties to those of the parent static stars, in order to provide more meaningful
averages. 

We conclude with a tribute to Pr. Michael Scholz, whose dynamical models of have been
a major contribution to our understanding of the relationships between atmospheric structure
and spectrophotometric properties of Mira variables. They remain precious today. 
\smallskip

\textit{Acknowledgments}. AL, PP \& AG thank the Programme National de Physique 
Stellaire (PNPS) and the Programme National Cosmologie \& Galaxies (PNCG), France, for 
recurrent support. AG is supported by the EU-FP7 programme through grant number 320360.

%
%
%
%

\end{document}